\documentclass[iop]{emulateapj}
\usepackage{xspace,graphics,graphicx}

\slugcomment{Accepted for publication in the Astronomical Journal}
\shorttitle{Spectroscopy of EU Cnc}
\shortauthors{Williams et al.}

\begin{document}
\title{Time-resolved Spectroscopy of the Polar EU Cancri in the Open
  Cluster Messier 67 \footnote{Some of the data presented herein were
    obtained at the W. M. Check Observatory, which is operated as a
    scientific partnership among the California Institute of
    Technology, the University of California and the National
    Aeronautics and Space Administration.  The Observatory was made
    possible by the generous financial support of the W.M. Keck
    Foundation.}}
    
\author{Kurtis A. Williams}
\affil{Department of Physics \& Astronomy \\ Texas A\&M University --
  Commerce \\ P.O. Box 3011, Commerce, TX, USA, 75429}
\email{kurtis.williams@tamuc.edu}

\author{Steve B. Howell}
\affil{NASA Ames Research Center \\ P.O. Box 1, M/S 244-30, Moffett Field, CA 94035}

\author{James Liebert}

\author{Paul S. Smith}
\affil{Steward Observatory \\ University of Arizona, Tucson, AZ}

\author{Andrea Bellini}
\affil{Space Telescope Science Institute \\ 3700 San Martin Drive, Baltimore, MD 21218}

\author{Kate H.~R. Rubin}
\affil{Max-Planck-Institut f\"ur Astronomie \\ K\"onigstuhl 17, 69117 Heidelberg, Germany}

\and
\author{Michael Bolte}
\affil{UCO/Lick Observatory \\ University of California\\ 1156 High St., Santa Cruz, CA, USA, 95064}

\begin{abstract}
We present time-resolved spectroscopic and polarimetric observations of the AM Her system EU Cnc.  EU Cnc is located near the core of the old open cluster Messier 67;  new proper motion measurements indicate that EU Cnc is indeed a member of the star cluster, this system therefore is useful to constrain the formation and evolution of magnetic cataclysmic variables. The spectra exhibit two-component emission features with independent radial velocity variations as well as time-variable cyclotron emission indicating a magnetic field strength of 41 MG. The period of the radial velocity and cyclotron hump variations are consistent with the previously-known photometric period, and the spectroscopic flux variations are consistent in amplitude with previous photometric amplitude measurements.  The secondary star is also detected in the spectrum.  We also present polarimetric imaging measurements of EU Cnc that show a clear detection of polarization, and the degree of polarization drops below our detection threshold at phases when the cyclotron emission features are fading or not evident.  The combined data are all consistent with the interpretation that EU Cnc is a low-state polar in the cluster Messier 67.  The mass function of the system gives an estimate of the accretor mass of $M_{\rm WD} \geq 0.68 M_\odot$ with $M_{\rm WD}\approx 0.83 M_\odot$ for an average inclination.  We are thus able to place a lower limit on the progenitor mass of the accreting WD of $\geq 1.43 M_\odot$.
\end{abstract}


\section{Introduction\label{sec.intro}}
Cataclysmic variables (CVs) are interacting binary systems in which a
white dwarf (WD) is accreting material from a low-mass companion star.  If
the WD has a sufficiently strong magnetic field, the
formation of an accretion disk is inhibited, and material accretes
directly onto one or more magnetic poles of the white dwarf.  These
binaries are known as AM Her systems or polars, after the high
fraction of polarized light detected in the systems.

The identification of CVs in star clusters provides interesting
constraints on the formation and evolution of the interacting system.
As all members of an cluster are coeval, the total age of the
system is known, as is the system's distance and metallicity.  Further, if the mass
and effective temperature of the white dwarf can be determined, then a
limit on the white dwarf's progenitor mass can be derived via the same
methods used to construct the initial-final mass relation \citep[e.g.,][]{2009ApJ...693..355W}.  Because
the white dwarf is likely re-heated to some extent by the ongoing
accretion, the constraint on the progenitor mass would be strictly a
lower limit.

CVs are quite common in globular clusters \citep[e.g.][]{1981ApJ...247L..89M,1995ApJ...455L..47G}.  The globular CV population tends to be centrally concentrated \citep[e.g.][]{1995ApJ...455L..47G,2001ApJ...563L..53G}.  This result is likely explained by the formation of tight binaries by stellar encounters in globular cluster cores \citep[e.g.][]{2002ApJ...565.1107P,2003ApJ...591L.131P}.  As such, globular cluster CVs are excellent tracers of a globular's dynamic history, but these CVs may not shed much light on the formation and evolution of CVs in the galactic field where stellar encounters are rare.

Open star clusters may therefore be a more useful laboratory for studying CV evolution.  The stellar densities are far lower than in globular clusters, even in the cluster core, and dynamical simulations suggest that CV formation is not enhanced by the stellar encounters that do occur \citep{2002ApJ...571..830S}, though a small fraction of open cluster CVs may still be formed by stellar exchanges \citep{2006ApJ...646..464S}.  Open clusters also span a wide range of ages, metallicities, and stellar masses, raising the potential to study how these parameters impact CV formation and evolution more precisely than possible from studies of field CVs.

Unfortunately, the number of CVs in open clusters is small, and none are well-studied.  In the ancient, dense, metal-rich open cluster \object{NGC 6791}, two spectroscopically-confirmed CVs are known \citep{1997ApJ...491..153K,2003AJ....125.3175M}; \citet{2007A&A...471..515D} identify a suspected third cluster CV based on photometric properties and conclude that all three CVs are likely cluster members.   One CV, \object{EU Cnc}, is known in the open cluster M67 and described in detail below.  These four objects are the only \emph{confirmed cluster member} CVs, and due to the large distances of the clusters [$(m-M)_V=13.4$ for NGC 6791 and $(m-M)_V=9.97$ for M67], spectroscopic studies of these CVs with the same precision and the same techniques (such as tomography) as current field CV studies require significant time on 8 m-class telescopes and larger. 

Other candidate open cluster CVs have been suggested: \citet{2004AJ....128..312M,2006AJ....131.1090M} identify a CV in the field of the $\sim 2-3$ Gyr-old open cluster \object{NGC 2158}, though it may lie foreground to the cluster.  One CV is identified photometrically in the field of the 3.5 Gyr-old cluster \object{NGC 6253}, but no membership information is available \citep{2010A&A...509A..17D}.  The rich open cluster \object{M37} (age $\sim 550$ Myr) has two CV candidates identified photometrically by \citet{2008ApJ...675.1254H}.  Finally, an X-ray source in the field of the cluster \object{NGC 6819} (age $\sim 2-2.4$ Gyr) has properties consistent with CVs, but its true nature and cluster membership are unconfirmed \citep{2012ApJ...745...57G}.  The \object{Hyades} contains at least one pre-CV, \object{V471 Tau} \citep[e.g][]{1972A&A....17..437V}, but no actively accreting systems.

\subsection{EU Cancri}
EU Cnc was detected as variable object in the field of the old open
star cluster Messier 67 by \citet{1991AJ....101..541G}, who identified
it as a likely AM Her system based on the similarities of its light
curve to that of VV Pup.  They determined EU Cnc has a photometric period of $2.091\pm 0.002$
hr with variations through a CuSO$_4$ ($U$ band) filter of 0.6 mag.
\citet{1993A&A...269..175B} detected an X-ray source coincident with the
optically-variable source; its very soft X-ray hardness ratio is
typical for AM Her systems in the ROSAT bands.
The AM Her nature of EU Cnc was confirmed by
\citet{1994A&A...290L..17P}, who obtained three 75-minute optical
spectra of the source.  These spectra exhibit cyclotron humps and
radial-velocity variable emission lines of H, \ion{He}{1},
\ion{He}{2}, and \ion{Fe}{2}.  Under the assumption that EU Cnc is
in the star cluster, \citet{1994A&A...290L..17P} conclude that the
absolute optical magnitude and X-ray luminosity of EU Cnc are
typical for low-state AM Her systems.

Subsequent X-ray studies by \citet{1998A&A...339..431B} using ROSAT
detected 100\% modulation of the soft X-ray flux with a period equal
to the optical period, indicative of accretion onto a single magnetic
pole.  Chandra observations by \citet{2004A&A...418..509V} detected
hard X-ray emission from EU Cnc, again typical for AM Her systems and
likely due to shocks in the accretion flow.

More recent time-series photometry from \citet{2005IBVS.5585....1N}
again detected high-amplitude optical modulation, with $V$-magnitudes
varying from 21.6 to 20.3 mag at the same 2.09 hr period of
\cite{1991AJ....101..541G}.  The variation amplitude was about 30\%
larger than in 1991, with that difference likely due to different
filters used in the two studies, $V$ containing a large cyclotron modulation. 

Based on the body of work on EU Cnc, \citet{2005IBVS.5585....1N} point
out an interesting conundrum.  The high-amplitude optical variability
is typical for AM Her systems in a high accretion state, while the
X-ray luminosity of EU Cnc, \emph{under the assumption it is a
  member of M67}, is typical of magnetic CVs in a
low accretion state. 

In this paper, we present time-resolved spectroscopy of EU Cnc
obtained serendipitously with the 10-m Keck telescope, as well as the
first polarimetric measurements of this system. After discussing the
observed phenomenology, we revisit the issue of EU Cnc's cluster
membership. We will show that the optical variability is almost entirely due to cyclotron
emission changing throughout the orbit and that the optical spectrum is typical of a low
accretion state polar. We detect the secondary star in the spectrum and, along with the 
other evidence, we show that EU Cnc is a member of the M67 open cluster.

\begin{figure*}
\begin{center}
\includegraphics[scale=0.67]{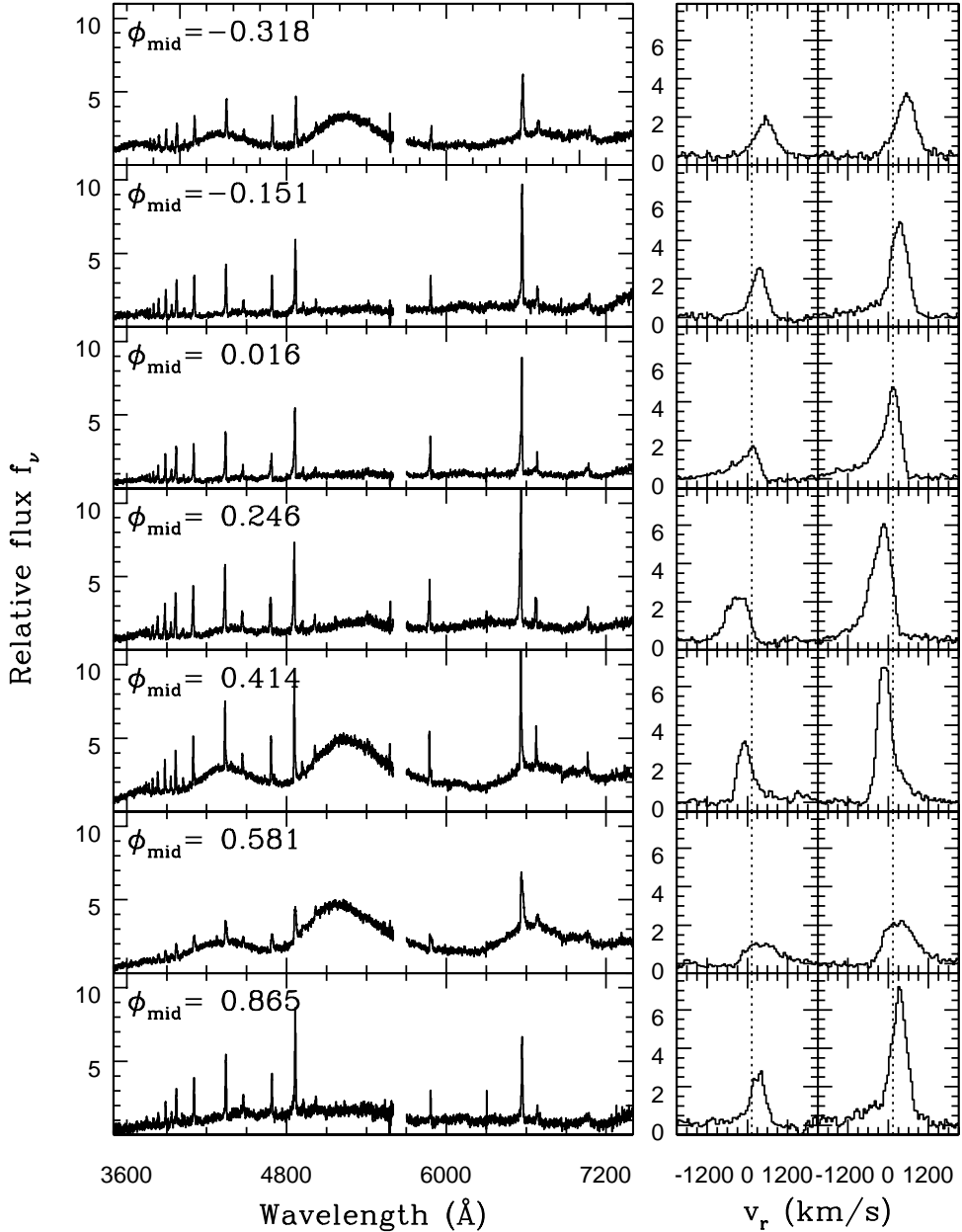}
\end{center}
\caption{Time-ordered spectroscopy of EU Cnc.  The left panels contain the entire spectrum; the gap at $\lambda\approx 5650$\AA\  is the gap between the blue and red arms of the spectrograph.  The right panels contain close-ups of the $\lambda$4686 \ion{He}{2} line (left) and H$\beta$ line (right); the radial velocity variations and variable line asymmetries are clearly visible.  Phase $\phi=0$ indicates the negative zero crossing of the narrow-line component of the emission lines.   \label{fig.allspec}}
\end{figure*}

\section{Time-Resolved Spectroscopy\label{sec.spec}} 
\subsection{Observations and Data Reduction\label{sec.spec.obs}}
We targeted EU Cnc serendipitously as part of a program to obtain
high signal-to-noise spectroscopy of WDs in Messier 67.
We obtained observations on UT 2007 Jan 19 with the low-resolution
imaging spectrometer (LRIS) on the Keck I telescope
\citep{1995PASP..107..375O,2004ApJ...604..534S}.  We obtained simultaneous
spectra with both blue and red sides of the spectrograph through a
multiobject slit mask with slitlet widths of 1\farcs 0. On the blue side,
we used the 400 lines mm$^{-1}$ grism blazed at 3400\AA\  for a
resulting spectral resolution of $\approx 7$\AA\  FWHM.  On the red side
of the spectrograph, we used the D560 dichroic and the 600 grooves
mm$^{-1}$ grating blazed at 7500\AA, for a resulting spectral
resolution of 4.8\AA\  FWHM.  The total spectral coverage ranged from
7400\AA\  blueward to the UV atmospheric cutoff.

The weather was nearly photometric through the entire observation, and
seeing was moderate at 0\farcs9 FWHM.  Seven exposures, each of
twenty-minute integration, were taken over a $\sim 2.5$ hr period,
with two short breaks for mask re-alignment. During the final
exposure, the flux dropped
dramatically for most stars, likely 
indicating that the mask was slightly misaligned.  

These observations were taken prior to the installation of the LRIS
atmospheric dispersion corrector, and due to constraints in slit mask
design, the slitlets were oriented nearly perpendicular to the
parallactic angle.  We attempted to minimize the effects of
atmospheric dispersion by using a blue filter in the guider camera,
but diminution of the UV light is severe at higher airmasses. 

We reduced the data using the \emph{onedspec} and \emph{twodspec}
packages in IRAF.  Overscan regions were used to determine and remove
the amplifier bias.  Flat-fielding on the blue side data was accomplished
using a piecewise-smooth response function as described in \citet
{2009ApJ...693..355W} to eliminate ringing due to a sharp inflection
point in the flat field at $\approx 4200$\AA.  Cosmic rays were
removed from each two-dimensional spectrum using the L.A.~Cosmic
Laplacian cosmic ray rejection routine \citep{2001PASP..113.1420V}.
Wavelength solutions on the blue side were derived from Hg, Cd, and Zn
arclamp spectra; on the red side, Ne and Ar arclamp spectra were
used.  These calibrations were obtained prior to the final mask
realignment.

Relative flux calibration was obtained using 1\arcsec-wide long-slit
spectroscopy of the spectrophotometric standards G191-B2B and G138-31
taken at parallactic angle near the start of the night.  No attempt
was made to obtain absolute spectrophotometric calibrations.

\subsection{Spectral Phenomenology\label{sec.spec.phenom}}
The time-series spectra are shown in Figure \ref{fig.allspec}.
Qualitatively, the spectra appear fairly typical for AM Her-type
systems in a low accretion state, similar to those observed for HU Aqr \citep{1994ApJ...424..967G}. 
Cyclotron humps are visible and variable in strength.
Emission lines of H and He are observed and have at least two
components, a narrow (unresolved) component and a broad component.
The emission lines have variable radial velocities, and the radial velocities of the two
line components are not in phase.  This type of emission line component behavior is often seen
in polars, for example VV Pup \citep{2008A&A...490..279M} whereby the narrow line component phases 
with the motion of the secondary star \citep{2008A&A...490..279M,2008AJ....136.2541H}.
The time scales for both the cyclotron
hump variations and emission line variations appear to be consistent
with the known photometric period, though additional time-series spectra would be required to prove these are identical.  We now quantify these phenomena.

We estimate the strength of the magnetic field using the spacing of the cyclotron humps in the optical spectra.  The locations of the peaks are estimated by fitting a 12th-order polynomial to the continuum, excluding the obvious emission lines.  We use equation 54 from \citet{2000PASP..112..873W} to get a magnetic field strength of $B\approx 41$ MG.  We do not see Zeeman split hydrogen absorption lines in the spectrum corresponding to this magnetic field strength, but later we will see that they are likely filled in by the red continuum contribution of the secondary star.

\subsubsection{Radial Velocity \label{sec.rv}}
Because these data were not obtained with the goal of obtaining precision radial velocities, and because the flexure in the spectrograph was significant over the course of the observations, it was not possible to obtain absolute radial velocities.  Attempts to use night sky emission lines as velocity zero points failed due to the lack of night sky lines in data from the blue arm of the spectrograph, where the majority of the CV's emission lines were located.

\begin{figure}
\begin{center}
\includegraphics[scale=0.4]{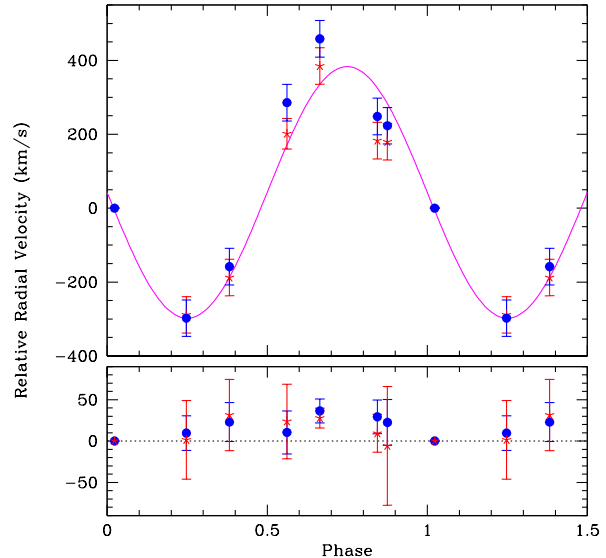}
\caption{Radial velocity curve for EU Cnc (top panel) and a G-type star in a neighboring slitlet (bottom panel).  Filled circles (blue in the online color version)indicate relative radial velocities from emission lines on the blue arm of the spectra; stars (red in the online version) indicate data from the red arm of the spectrograph.  The solid curve (magenta in the online version) is the best-fitting sine wave.\label{fig.rv}}
\end{center}
\end{figure}

Instead, relative radial velocities were obtained as follows.  First, the continuum was fit with a high-order polynomial and removed from each spectrum.  The radial velocity of the narrow component of the emission lines was determined by autocorrelation of each spectrum with that of an observation with an exposure midpoint close to zero phase.   This determination involved some iterative bootstrapping with the radial velocity fitting described further below.  The resulting radial velocities (relative to the observations closest to zero phase) are given in Table \ref{tab.vel} and shown in Figure \ref{fig.rv}.  The radial velocities at $\phi\approx 0.59$ are likely measuring different motion than the other points, as the narrow component of the emission lines disappears during these observations.

As a cross-check, we used the same method to determine the radial velocity of a G-type star targeted in a neighboring slitlet.  The radial velocity of this star should be constant over the set of observations, and although a slight positive velocity offset may exist, it is of low amplitude compared with the variations observed in the CV.

\begin{deluxetable*}{cclcccc}
\tablecolumns{7}
\tablewidth{0pt}
\tablecaption{Radial velocities of EU Cnc and an unrelated neighboring G-type star \label{tab.vel}}
\tablehead{\colhead{Obs. Midpt} & \colhead{Phase} & \colhead{Spectrograph}
  & \colhead{$v_{\rm EU\, Cnc, rel}$} & \colhead{$\sigma (v_{\rm EU\, Cnc, rel})$} & 
\colhead{$v_{\rm *,rel}$} & \colhead{$\sigma(v_{\rm *,rel})$} \\
(MJD) & & Arm & (km s$^{-1}$) & (km s$^{-1}$) & (km s$^{-1}$)  & (km s$^{-1}$) }
\startdata
54119.51236 & 0.682 &  Red &  384.9 & 13.8 & 27.8 &  5.2\\
54119.52732 & 0.854 &  Red &  183.0 & 23.2 &  9.4 & 20.8\\
54119.54231 & 0.026 &  Red &    0.0 &  0.0 &  0.0 &  0.0\\
54119.56151 & 0.246 &  Red &$-288.7$& 27.6 &  1.5 & 46.9\\
54119.57651 & 0.418 &  Red &$-187.5$& 28.1 & 31.5 & 42.4\\
54119.59146 & 0.590 &  Red &  201.4 & 25.6 & 23.8 & 44.3\\
54119.61542 & 0.865 &  Red &  179.7 & 23.5 &$-5.9$& 71.1\\
54119.51239 & 0.682 & Blue &  458.5 & 17.9 & 36.3 & 11.8\\
54119.52689 & 0.849 & Blue &  248.3 & 27.5 & 29.3 & 18.8\\
54119.54145 & 0.016 & Blue &    0.0 &  0.0 &  0.0 &  0.0\\
54119.56153 & 0.246 & Blue &$-297.7$& 30.5 &  9.7 & 19.9\\
54119.57613 & 0.414 & Blue &$-158.2$& 24.9 & 22.9 & 21.2\\
54119.59067 & 0.581 & Blue &  285.7 & 33.6 & 10.5 & 22.7\\
54119.61545 & 0.865 & Blue &  223.2 & 35.2 & 22.6 & 25.0\\
\enddata
\tablecomments{
  Phase $\phi=0$ indicates the negative zero
  crossing of the narrow-line component of the emission lines; velocities are relative to the exposure closeest to $\phi=0$.}
\end{deluxetable*}

The phase and amplitude of the radial velocity variations were determined by fitting a function of the form
$$v_{rel}=K\sin[2\pi(\phi-\phi_0+0.5)]+v_0$$ 
to the radial velocity data.  The period was fixed to the known optical period of 2.091 hours, and the phase shift $\phi_0$, amplitude $K$, and the relative velocity zeropoint $v_0$ were allowed to vary.  The best-fit values are $K=340\pm 20\, {\rm km\;s^{-1}}$ and $v_0=+43\pm 37$ km s$^{-1}$.  We emphasize that this velocity zero point is a velocity relative to the spectra used as our zero velocity, which were selected since they are the closest data to a phase of zero.  Based on the detailed observation of narrow line components in the polars VV Pup \citep{2008A&A...490..279M} and EF Eri \citep{2008AJ....136.2541H}, mapping to the motion of the secondary star, the typical orbital phase 0.0 used for cataclysmic variables would occur near phase 0.5 as shown in Figures \ref{fig.allspec} and \ref{fig.rv}. However, we do not have sufficient data here (e.g., a velocity curve of the photosphere of the secondary star) to state this fact with absolute certainty.  As these velocities are not absolute radial velocities, they cannot be used for cluster membership determination or rejection. 

Qualitatively, the sine curve is not a superb fit to the data, especially for phases between 0.5 and 1.  We note that, at these phases, the narrow component of the lines is relatively weak and the broad component strong.  In fact, at phase $\phi=0.59$, the narrow component of the lines is not visible.  These velocity measurements are likely non-Keplerian streaming motion.  However, the excellent fit for the points with phases between 0 and 0.5, where the emission lines are dominated by the narrow component, suggests that our fit  amplitude and phase shift are not unreasonable.

\subsubsection{Photometric Variability}

As mentioned above, absolute spectrophotometry was not a goal of our
observations; slit losses as a function of wavelength could be
significant and time-variable due to both seeing and atmospheric
dispersion effects.  Even so, it would be instructive to compare photometry
derived from our spectroscopy with previous photometric monitoring of
this system.  We therefore calculate broad-band photometry
by folding our (relative) flux-calibrated spectra through filter
response functions.  Due to the significant loss of UV light in the
later exposures, we restrict our analysis to the $g$-band. 

In order to correct for slit losses and variable atmospheric
absorption, we perform the same calculation for a K-type star targeted
in a different slitlet on the same mask (note that this is not the same star used for radial velocity comparisons in \S\ref{sec.rv}).  The star is located at
$\alpha({\mathrm J2000})=8^{\mathrm h} 51^{\mathrm m} 35\fs46$,
$\delta({\mathrm J2000}) =11\degr 50\arcmin 19\farcs1$, and, in our
photometry (Williams et al, in preparation), has $g=20.324\pm 0.032$.  This star also
appears in SDSS DR7, with PSF magnitude $g=20.263\pm 0.038$.  Due to
the higher precision of our data, we adopt our photometry for this
star.

For each exposure, the calibrated spectra of both EU Cnc and the
comparison star are folded through the $g$-band filter response
function using the Synphot synthetic photometry package in STSDAS.
All magnitudes are calculated as AB magnitudes.  Zero-point offsets in
$g$ are calculated from the comparison star; these offsets are
then applied to the calculated magnitudes for EU Cnc.  The results are
included in Table \ref{tab.phot}.  The errors in this photometry
are uncertain.  The random errors due to photon shot noise in the
source and sky are small, $\leq 0.003$ mag.  However, systematic
errors such as differential slit losses likely dominate the
photometric uncertainty.  The magnitude of the error in absolute
photometry is at least $0.03$ mag (the error in the broadband
photometry of the comparison star).

\begin{deluxetable}{cccc}
\tablewidth{0pt}
\tablecolumns{4}
\tablecaption{Time Series Broad-band Photometry of EU
  Cnc\label{tab.phot}}
\tablehead{\colhead{Midpoint Obs.} & \colhead{$g_{*,{\rm meas}}$
\tablenotemark{a,b}} & \colhead{$g_{\rm EU\,Cnc,\, meas}$\tablenotemark{a}} &
\colhead{$g_{\rm EU\, Cnc,\, corr}$\tablenotemark{c}}\\ 
(HJD) & (mag) & (mag) & (mag)}
\startdata
2454120.01789 & 20.563 & 20.806 & 20.567\\
2454120.03239 & 20.676 & 21.601 & 21.249\\
2454120.04695 & 20.480 & 21.672 & 21.516\\
2454120.06703 & 20.703 & 21.266 & 20.887\\
2454120.08163 & 20.614 & 20.440 & 20.150\\
2454120.09617 & 20.467 & 20.429 & 20.286\\
2454120.12095 & 21.147 & 21.706 & 20.883\\
 \enddata
 \tablenotetext{a}{Measured flux folded through $g$ filter response}
 \tablenotetext{b}{Instrumental magnitude of the neighboring K star}
 \tablenotetext{c}{EU Cnc photometry corrected to standard system}
\end{deluxetable}

We compare our derived photometry with the time-series photometry of
\citet{2005IBVS.5585....1N}.  These published data were taken in $V$;
we converted these magnitudes to $g$ using the Population I
transformation equations in Table 4 of \citet{2006A&A...460..339J} and
the \bv$=0.41$ color of EU Cnc from \citet{1991AJ....101..541G}. As
these transformations are for single stars, and as the color of EU Cnc
is likely changing as a function of phase, we emphasize that these
transformations are meant to be illustrative only.  Since the ephemeris
of EU Cnc is not sufficiently well-determined to allow us to phase the data
precisely, we added an arbitrary phase shift to the published
photometry.

\begin{figure}
\begin{center}
\includegraphics[scale=0.4]{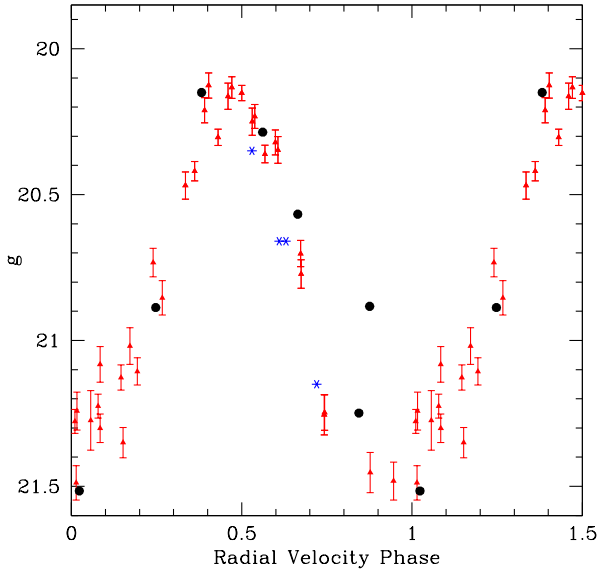}
\caption{The light curve for EU Cnc.  Large filled circles are the
  spectrophotometry data
  from this work; error bars are not included but are likely at least
  0.03 mag. Filled triangles with error bars (red in the online version) are data from
  \citet{2005IBVS.5585....1N}, shifted in phase by an arbitrary
  offset.  Stars (blue in the online version) are spectropolarimetric data form this work.
  The light curves are qualitatively nearly identical between these
  data sets, suggesting little change in the state of the polar
  between early 2004 and late 2007. \label{fig.lc}}
\end{center}
\end{figure}

The results of this comparison are shown in Figure \ref{fig.lc}.  With
the exception of our observation at $\phi=0.865$, our corrected
spectrophotometry and the published time-series photometry agree very
well in both shape and amplitude.  

\subsection{Time-series Polarimetry}
Circular polarimetry of EU Cnc was obtained with the SPOL
spectropolarimeter \citep{1992ApJ...398L..57S} on the 2.3-m Bok
Telescope at Steward Observatory in 2007 December.  The observations
were taken in the imaging mode of the instrument using a Hoya HA30$+$Y48 filter combination,
giving a broad bandpass of $\approx 4800-7000$ \AA.  Image acquisition and data reduction
follow those described in Smith et al. (2002), but modified as appropriate
for circular polarimetry.  A $\lambda$/4 wave plate is used to convert
incident circular polarization to linear and the Wollaston prism separates
the light into two orthogonally polarized beams that are focused onto a
CCD.  Two separate reads of the CCD are made with the wave plate rotated
through four positions to determine the circular V Stokes parameter.  Two
consecutive 10-minute exposures (150 s per wave plate position) were
obtained on each of two nights.  The sky was clear during the observations, but no
attempt at absolute calibration was made because of the non-standard
filter bandpass used.  Photometric information was extracted using a
circular aperture of radius 6\arcsec. 

The observing log and resulting circular polarization measurements 
($V/I = v$) are given in Table \ref{tab.polarimetry}.  Also given are
the background subtracted counts and the phase of the midpoint of each
observation relative to first exposure on the first night assuming a photometric
period of 2.091 h.  There is a significant detection of polarization
in the first exposure, with $v=-12.7\pm 1.8\%$.  The degree of
polarization and the total flux both drop significantly in the second
exposure ($v=-6.3\pm 2.3\%$).  The third exposure, taken two nights
later, is at nearly the same phase as the second exposure.  The total
flux is slightly lower and there is no significant detection of
polarization ($v=-2.8\pm 2.6$); the final exposure is the faintest of
the four and has no significant polarization.

\begin{deluxetable*}{cccccc}
\tablecolumns{6}
\tablewidth{0pt}
\tablecaption{Polarimetric Observations of EU Cnc\label{tab.polarimetry}}
\tablehead{\colhead{Observation Date} & \colhead{Obs.\ Time} &
  \colhead{Relative Phase\tablenotemark{a}} & \colhead{Counts} &
  \colhead{$V/I$} & \colhead{$\sigma(V/I)$} \\ (UT) & (UT) & ($\Delta
  \Phi$) & (ADU) & (\%) & (\%)}
\startdata
2007 December 14 & 10:28 & \nodata & 74147 & $-12.7$ & 1.8 \\
2007 December 14 & 10:39 & 0.08 & 55485 & $-6.3$ & 2.3 \\
2007 December 16 & 08:41 & 0.10 & 48303 & $-2.8$ & 2.6 \\
2007 December 16 & 08:52 & 0.19 & 30698 & 4.6 & 4.0 \\
\enddata
\tablenotetext{a}{Phase relative to first polarimetric observation}
\end{deluxetable*}

As these observations were taken 11 months after the spectroscopy, and
as the ephemeris of EU Cnc is not sufficiently well known, the
spectroscopic phase of these observations cannot be calculated -- the accumulated uncertainty in phase given the 0.002 hr uncertainty in the photometric period of \citet{1991AJ....101..541G} is $\approx 7.5$ cycles.
However, from the total counts in the polarimetry measurements, we
know that these observations were taken on the declining portion of
the light curve.  We therefore calculate relative magnitudes from the
total intensity measurements in Table \ref{tab.polarimetry} and add an
arbitrary magnitude zero point and phase shift to place these points
in the folded light curve of Figure \ref{fig.lc}.  

From this exercise,
we see that the rate of decline in the observed intensity agrees with
that observed in the time-series photometry of
\cite{2005IBVS.5585....1N} and in our spectrophotometry, indicating that
the spectroscopic phase of the first polarimetric observations was 
$\phi \approx 0.5\mathrm{~to~} 0.65$. Comparison with the time-series spectra in
Figure \ref{fig.allspec} shows that this corresponds to a phase when
the cyclotron humps are dominant in the spectrum; these humps then
vanish by a phase $\phi=0.85$.  This is consistent with the strong
degree of polarization observed in the first exposure, and the
weaker/insignificant polarization in the other polarimetry exposures.

In summary, we detect significant polarization from EU Cnc at a phase
likely corresponding to strong spectroscopic cyclotron emission
features, and the degree of polarization decreased below our detection
threshold at phases when the cyclotron features were fading or not
evident in the spectrum.  This proves that the features are indeed
cyclotron emission.  From a semantic point of view, this detection of
significant polarization also confirms that we can use the moniker
``polar'' to refer to EU Cnc.

\section{Discussion\label{sec.disc}}
\subsection{Cluster Membership\label{sec.disc.member}}

EU Cnc is projected $\approx 1\farcm 7$ away from the cluster center
defined by \citet{1993AJ....106..181M} (0.4 pc at the cluster
distance), well within the observed core radius of $\approx 4\farcm 5$
\citep[e.g.,][]{1976AJ.....81..835T,2003A&A...405..525B}.   However, this proximity alone is not sufficient evidence of cluster membership.

\begin{figure}
\begin{center}
\includegraphics[scale=0.4]{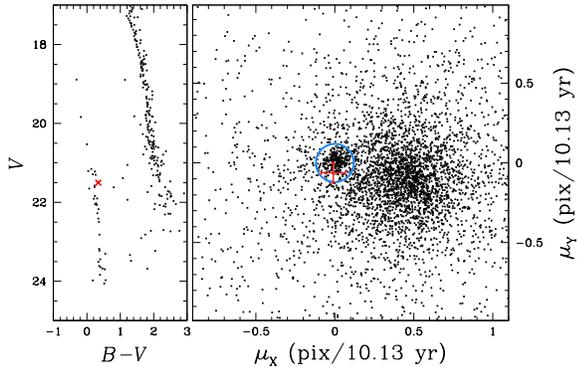}
\end{center}
\caption{The proper-motion selected color-magnitude diagram (CMD, left) for M67 and the vector point diagram (VPD, right) for all stars in the field as determined by \citet{2010A&A...513A..50B}.  The cross (red in the online version) in the CMD indicates the location of EU Cnc.  The circle (blue in the online version) in the VPD indicates the selection criteria Bellini et al.~ used to select cluster members; the error bars (red in the online version) show the location and uncertainty in the proper motion of EU Cnc.  The qualitative data are convincing that EU Cnc is a proper-motion member of the cluster. \label{fig.cv_pm}}
\end{figure}

\citet{2010A&A...513A..50B} present proper-motion data for white dwarfs in M67, precise to $V\approx 26$, obtained from multiple-epoch imaging from the Canada-France-Hawaii Telescope and the Large Binocular Telescope.  Figure~\ref{fig.cv_pm} shows the vector-point diagram of objects within $20^\prime$ from the cluster center (right panel). The circle indicates the proper motion membership cut adopted by \citet{2010A&A...513A..50B}; the error bars show the displacement uncertainty for EU Cnc, solidly inside the membership circle. In addition, the proper-motion selected color magnitude diagram (left panel) shows that EU Cnc lies on the WD cooling sequence of M67.

Astrometry performed on EU Cnc and 3 encircling brighter known member stars, using the POSS I and II plates as well as a deep, $0\,\farcs4$ seeing image obtained by us at the Kitt Peak WIYN telescope in 2007, show that the polar's location relative to the other three stars changes by $0\,\farcs3 \pm 0\,\farcs5$ over the 60 year interval. This is a similar uncertainty to that for each of the brighter stars relative to each other. 
 We therefore feel confident that EU Cnc is a member of the M67 star cluster.

\subsection{A low-state polar in M67}

\citet{2005IBVS.5585....1N}, suggest that the luminosity of EU Cnc is lower than other high state polars and fully consistent with the luminosity of low-state polars in globular clusters; the distances to field magnetic CVs are too uncertain to allow for such a precise comparison. The spectra obtained in this study, as well as those prior, show the clear indications of a low state polar: a steep Balmer decrement (compared to a flat decrement and Balmer jump in emission in high states) and the separable narrow and broad emission line components. \citet{1999ASPC..157...63W} examine the high-low state range of polars and
notes that the magnitude difference is typically near 3-4 for a 2 hour polar. Additionally, low state polars have absolute magnitudes near 11-12 in $V$ compared with 8-9 in $V$ when in a high state. If EU Cnc were in a high state with its observed apparent $V$ magnitude, it would need to reside at a distance of near 3100 pc. However, in its low state and $V\!\sim$21, we find it to be at a distance of approximately 850 pc or the same distance as M67 \citep{2009ApJ...698.1872S}.

\begin{figure}
\begin{center}
\includegraphics[scale=0.35,angle=270]{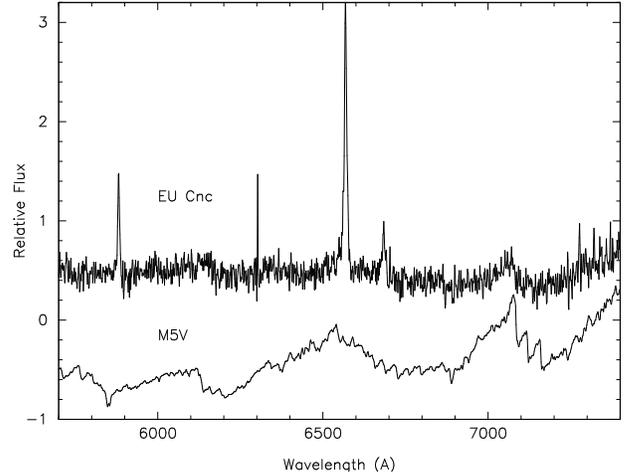}
\end{center}
\caption{Red spectrum from phase 0.865 in Fig. 1 along with a scaled M5V spectrum taken from the Jacoby Atlas \citep[star 57,][]{1984ApJS...56..257J}. The M5V star is scaled 1.4 magnitudes fainter than the stellar continuum, approximating the level of contribution is provides to the red end of the spectrum of EU Cnc.
\label{fig.m5v}}
\end{figure}

Additionally, we detect the secondary star in the optical spectrum of EU Cnc (see the late phase red spectra in Fig.~\ref{fig.allspec} and Fig.~\ref{fig.m5v}) by noting the TiO absorption bands
chopping into the spectrum at the characteristic locations near 6800\AA ~and 7100\AA. The shape and relative amplitude of the TiO humps are a good match to a M5V star which is also the expected spectral type of the secondary star in EU Cnc (see below). Noting the dilution of the TiO bands compared to a single M5V star, we determine that the secondary star contributes 20-25\% of the continuum redward of 6000\AA.
This estimate suggests that the  M5V secondary star will have an apparent magnitude near 22.5 and, for an its M$_V$=12.6, yields a distance to EU Cnc of 832 pc, again the same as M67. The M star continuum is also the likely cause of filling in the Zeeman spilt H$\alpha$ absorption components. 

\subsection{The masses of the white dwarf accretor and its progenitor star}

An advantage of finding a magnetic CV in an open star cluster is that numerous constraints can be placed on the progenitor system.   As EU Cnc is a member of Messier 67, its initial metallicity and total evolutionary age are identical to the cluster's characteristics. 
With some reasonable assumptions, we can go one step further to constrain the white dwarf's mass and cooling age, which we can then use to constrain the white dwarf progenitor's mass.  We describe these steps in detail below, but note that this methodology has been used in numerous open cluster and field star studies to determine WD progenitor masses \citep[e.g.][]{2005ApJ...630L..69L,2008ApJ...676..594K,2008A&A...477..213C,2009ApJ...693..355W,2012MNRAS.423.2815D}

WD masses are often determined through model atmosphere fits of the WD spectrum, but absorption lines from the white dwarf photosphere are not convincingly evident in our data, being filled in and misshaped by the emission.  However, we note that \citet{1985ASSL..113..151L} and \citet{2008A&A...490..279M} determined that the ``center of mass" for the narrow emission lines originates between $L_1$ and the center of mass of the donor, that is they approximate the center of mass of the secondary star. We therefore assume that the narrow component of the emission lines originates from the center of mass of the secondary star and use our velocity amplitude to determine the mass function of the system and to estimate the mass of the white dwarf, $M_{\rm WD}$.

The mass function $f(M)$ of the system is
\begin{equation}
f(M)=\frac{(M_{\rm WD}\sin i)^3}{(M_{\rm WD}+M_2)^2}=\frac{PK^3 (1-e^2)^{3/2}}{2\pi G}
\end{equation}
where $M_2$ is the mass of the donor star.  Assuming a circular orbit ($e=0$) with a period $P=2.091$ hr and the velocity amplitude $K=340$ km s$^{-1}$, from Eq. 1 we find $f(M)=0.356 M_\odot$.

Since we do not see eclipses, we can constrain the inclination to be $\lesssim 74\degr$ \citep[e.g.][]{1995cvs..book.....W}.  If we assume $M_{\rm WD}>>M_2$, we therefore find that $M_{\rm WD}\geq 0.40 M_\odot$.  If instead we  assume an average value of $<\sin^3 i>=0.679$ \citep[e.g.][]{1974AJ.....79..967T}, then we find $M_{\rm WD}=0.52 M_\odot$.   

However, the mass of the donor star is not negligible.  \citet{2001ApJ...550..897H} present detailed evolution models for the secondary stars in cataclysmic variables and note that for those systems with orbital periods below the period gap ($< 2.5 \mathrm{hr}$), the secondary stars follow the normal main sequence mass-radius relationship. We therefore expect the mass of the secondary star in EU Cnc to be  $M_{\rm 2}=0.21 M_\odot$ with a radius of  $R_{\rm 2}=0.22 R_\odot$. These value are roughly those of an M5V star.  Adopting this mass for the secondary star, we find $M_{\rm WD}\gtrsim 0.68\,M_\odot$ (since $i\lesssim 74\degr$) and $M_{\rm WD}=0.83\,M_\odot$ for $<\sin^3 i>=0.679$.

We note that these masses assume that our velocity amplitude is correct.  As noted in \S\ref{sec.rv}, our value of $K$ assumes that our velocity amplitude is resolving the narrow component of the emission lines and that the narrow lines trace the center of mass of the secondary star.  Higher spectral resolutions would be necessary to test the first assumption, and so we do not have good constraints on the errors or our confidence limits on the white dwarf mass.  However, the WD mass estimates are similar to or higher than the spectroscopic masses of young member WDs in M67 ($M = 0.5\, M_\odot$ to $0.6\, M_\odot$, Williams et al. in preparation).  Further, through a lucky coincidence, our ignorance of the WD mass does not significantly impact estimates of the WD's progenitor mass.  We therefore proceed (perhaps quixotically) to constrain the WD's progenitor mass.

To constrain the progenitor mass, we use the methodology of \citet{2009ApJ...693..355W}.  To summarize, we use the effective temperature and mass of the WD to determine its cooling age (the elapsed time since the WD emerged from the AGB progenitor star).  This cooling age is subtracted from the cluster age to get the progenitor star's nuclear lifetime; stellar evolutionary models are then used to infer the progenitor star mass.

The WD mass estimates are sufficiently high to conclude that the WD has a carbon-oxygen core, indicating that the common envelope phase for the progenitor system did not occur before helium ignition in the WD progenitor.  Since the AGB phase of stellar evolution is relatively short compared to the WD cooling times and progenitor nuclear lifetimes we estimate for this star, we assume that any effect of the common envelope phase on the calculation of the progenitor nuclear lifetime is minimal.  

To estimate the effective temperature of the WD, we assume that the minimum luminosity of the system is  due to solely to the combined light of the M5V secondary and the WD photosphere, and we assume that the WD has not been heated by the ongoing accretion.  Under these assumptions, the WD has $M_V=12.5$.  We then interpolate WD photometric evolutionary tables provided online by P.~Bergeron\footnote{\url{http://www.astro.umontreal.ca/\~{}bergeron/CoolingModels}} and computed from color and model calculations of \citet{2006AJ....132.1221H}, \citet{2006ApJ...651L.137K}, \citet{2011ApJ...730..128T}, and \citet{2011ApJ...737...28B} to find the WD's cooling age as a function of mass at which its luminosity is equal to the observed minimum luminosity of the CV.  Despite the uncertainty in the WD mass, the cooling age of the WD is relatively insensitive to its mass, at least over the range of 0.5$M_\odot$ to 1.0$M_\odot$, at $\log \tau_{\rm cool}=8.86$ to 8.95.   If the WD has experienced significant heating, or if the minimum luminosity is significantly contaminated by light from the accretion stream and/or the accretion column, then its actual cooling age will be older than these calculations.  This means that our derived progenitor mass is a \emph{lower limit} on the WD's actual progenitor mass.

We now calculate the WD progenitor mass.  We adopt an age of M67 of 3.5 Gyr to 4.5 Gyr, and a cluster metallicity of $Z_{M67}\approx Z_\odot$ and the calculated primordial value of $Z_\odot=0.0142$ \citep{2010Ap&SS.328..179G}.  Since the lower limit on the WD cooling age is well-constrained, we obtain a fairly precise lower limit on the progenitor mass of  $M_{\rm WD,init}=1.43 M_\odot \pm 0.07 M_\odot$, with the majority of the uncertainty due to the uncertainty in the age of M67.   For comparison, the current main sequence turnoff mass (using the same models and inputs) is $1.35 M_\odot$.  

If we relax the assumption of no heating of the WD, then the WD's cooling age (time since its emergence from the red giant progenitor) could be significantly longer; this would imply a larger progenitor mass.  Therefore, we can only constrain the WD progenitor mass calculated above to be a lower limit; i.e. $M_{\rm WD,init}\geq 1.43 M_\odot$.

\section{Conclusions}
We have presented new observations of the polar EU Cnc.  In particular, proper motion studies strongly indicate that this system is a bona-fide member of the old open star cluster M67, which provides strong constraints on EU Cnc's total age, metallicity, and progenitor mass.  Using phase-resolved optical spectra and polarimetry, we have shown that EU Cnc exhibits all the properties of a low state polar. Both assumed ``best fit" polar parameters as well as standard values for the donor star lead to conclusions that are consistent with the system's membership in M67.  EU Cnc is one of only four CVs confirmed to reside in open clusters and the only confirmed magnetic CV in an open cluster.  More detailed photometric, polarimetric, and spectroscopic observations on large telescopes will be required to refine the system parameters and model the system with state-of-the-art analyses. However models of magnetic cataclysmic variable formation and evolution are not usually attempted and never well constrained; further study of EU Cnc will greatly aid in these endeavors. 

\acknowledgements
K.A.W.~ is grateful for the financial support of National Science
Foundation awards AST-0397492 and AST-0602288.  The authors thank S.~Kafka for discussions on this system and for providing the original data from \citet{2005IBVS.5585....1N}.  We also thank G.~Schmidt for providing time from his observing run to obtain the polarimetry in this paper, D.~ Wickramasinghe for assistance in measuring the magnetic field strength, and the anonymous referee for suggestions leading to improvement of this paper.
The authors wish to recognize and acknowledge the very significant
cultural role and reverence that the summit of Mauna Kea has always
had within the indigenous Hawaiian community.  We are most fortunate
to have the opportunity to conduct observations from this mountain.

{\it Facilities:} \facility{Keck:I (LRIS-B)}, \facility{Bok (SPOL)}

\bibliographystyle{apj}

\end{document}